\documentclass[twocolumn,showpacs,aps,prl,superscriptaddress]{revtex4}

\usepackage{dcolumn}
\usepackage{amsmath}
\usepackage{epsfig}

\input pubboard/babarsym
\def\Dp      {\ensuremath{D^+}}
\def\Dm      {\ensuremath{D^-}}
\def\Dstar   {\ensuremath{D^{*}}}
\def\Dstarp  {\ensuremath{D^{*+}}}
\def\Dstarm  {\ensuremath{D^{*-}}}

\newcommand{\etal}{{\it et al.}}

\newcommand{\BABARPubYear}    {01}
\newcommand{\BABARPubNumber}  {16}

\newcommand{\SLACPubNumber} {8931}
\newcommand{\LANLNumber} {0107068}

\def\figurebox#1#2#3{%
    \def\arg{#3}%
    \ifx\arg\empty
    {\hfill\vbox{\hsize#2\hrule\hbox to #2{\vrule\hfill\vbox to #1{\hsize#2\vfill}\vrule}\hrule}\hfill}%
    \else
    {\hfill\epsfbox{#3}\hfill}%
    \fi}

\long\def\inst#1{\par\nobreak\kern 4pt\nobreak
    {\it #1}\par\vskip 10pt plus 3pt minus 3pt}

\begin{document}

\begin{flushleft}
\babar-PUB-\BABARPubYear/\BABARPubNumber \\
SLAC-PUB-\SLACPubNumber \\
hep-ex/\LANLNumber \\[10mm]

\end{flushleft}

\title{ 
\vskip 10mm
{\Large \bf\boldmath Search for the Decay $B^0 \rightarrow \gamma \gamma$}
}

%
\author{B.~Aubert}
\author{D.~Boutigny}
\author{J.-M.~Gaillard}
\author{A.~Hicheur}
\author{Y.~Karyotakis}
\author{J.~P.~Lees}
\author{P.~Robbe}
\author{V.~Tisserand}
\affiliation{Laboratoire de Physique des Particules, F-74941 Annecy-le-Vieux, France }
\author{A.~Palano}
\affiliation{Universit\`a di Bari, Dipartimento di Fisica and INFN, I-70126 Bari, Italy }
\author{G.~P.~Chen}
\author{J.~C.~Chen}
\author{N.~D.~Qi}
\author{G.~Rong}
\author{P.~Wang}
\author{Y.~S.~Zhu}
\affiliation{Institute of High Energy Physics, Beijing 100039, China }
% ...
\collaboration{The \babar\ Collaboration}
\noaffiliation

\date{\today}

\begin{abstract}
We present a limit on the branching fraction for the decay $B^0 \rightarrow \gamma \gamma$
using data collected at the $\Upsilon(4S)$ resonance with the \babar\ detector at the PEP-II 
asymmetric--energy $e^+e^-$ collider.
Based on the observation of one event in the signal region, out of a sample of 
21.3 $\times$ 10$^6$ $e^+e^- \rightarrow \Upsilon(4S) \rightarrow B\overline B$ 
decays, we establish an upper limit on the branching fraction of
${\cal B}(B^0 \rightarrow \gamma \gamma ) < 1.7 \times 10^{-6}$
at the 90\% confidence level. 
This result substantially improves upon existing limits.
\end{abstract}


\maketitle

\begin{figure}[ttt]
\begin{center}
\epsfig{file=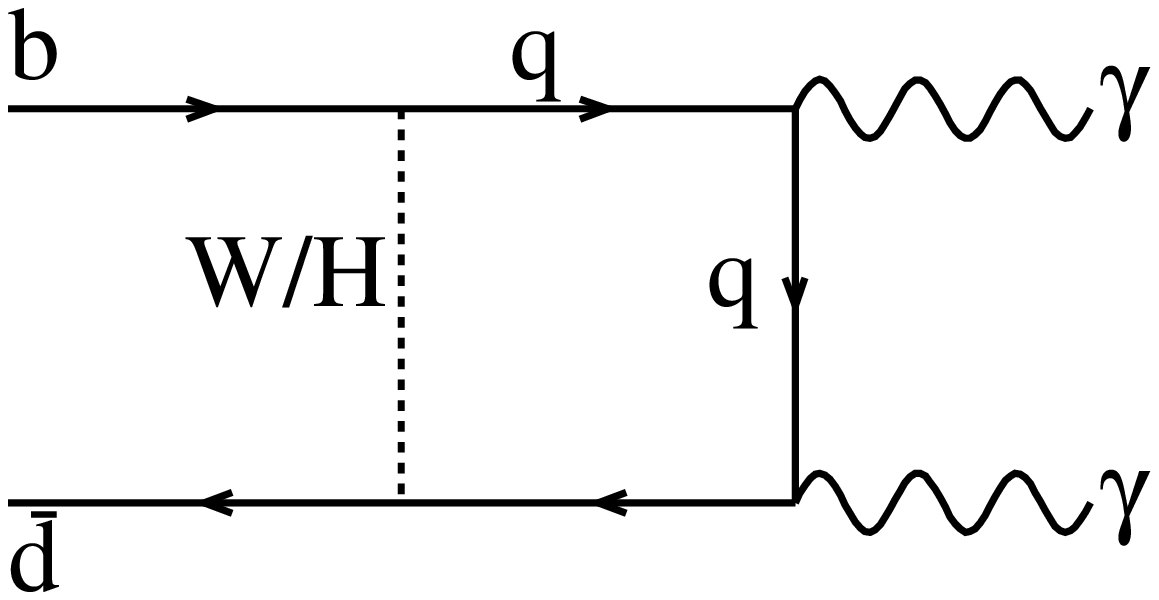,width=0.46\linewidth}\hfill
\epsfig{file=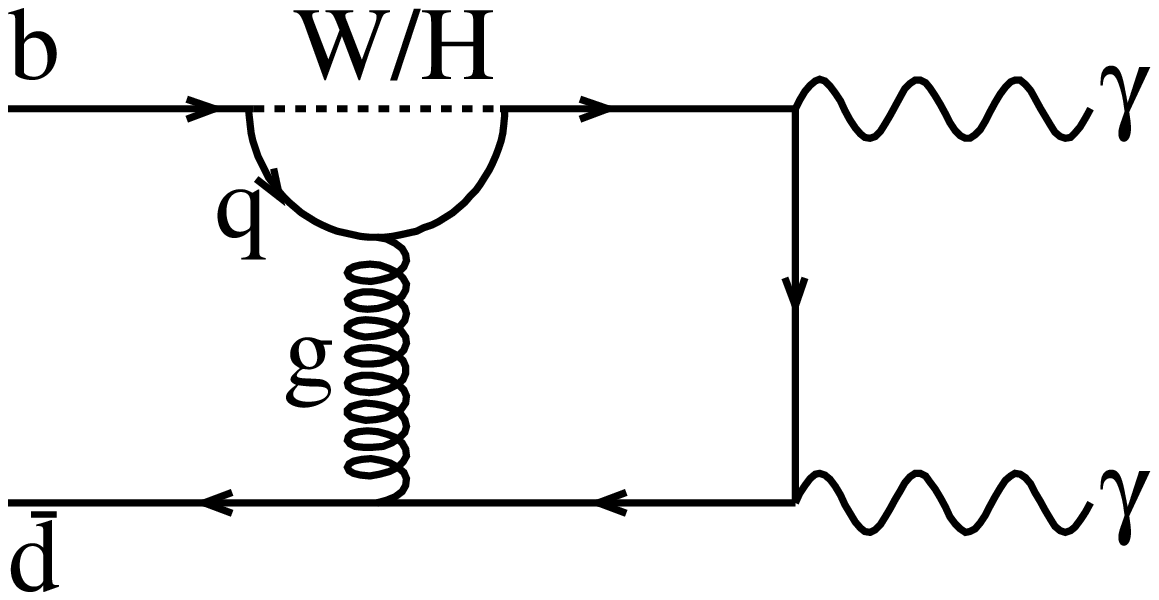,width=0.46\linewidth}
\end{center}
\caption
{Examples of possible diagrams responsible for 
the decay $B^0 \rightarrow \gamma \gamma$. In these diagrams
q = u, c, or t, and  H is a hypothetical charged non-Standard Model Higgs boson.
\label{fig:feynman}}
\end{figure}

%
%
In the Standard Model the decay  $B^0 \rightarrow \gamma \gamma$ proceeds via a
second-order weak transition, including gluonic penguins, followed by annihilation 
(Fig.~\ref{fig:feynman}).
Standard Model predictions for the branching fraction of 
these effective flavor--changing weak neutral current processes
range from 0.1 to $2.3\times 10^{-8}$ \cite{b0gammagammaSUSY}.

Physics beyond the Standard Model can enhance this branching fraction  
by as much as two orders of magnitude, particularly in the case of 
two--Higgs models \cite{b0gammagamma2Higgs}.
Other particles from the supersymmetric spectrum can further modify the Standard Model
expectation \cite{b0gammagammaSUSY}.
The current best limit on the branching fraction for $B^0 \rightarrow \gamma \gamma$,
from the L3 experiment \cite{L3} at the CERN LEP collider, is
${\cal B}(B^0 \rightarrow \gamma \gamma ) < 3.9 \times 10^{-5}$ (90\% confidence level).

%
%
In this Letter we present an analysis based on 
data taken with the \babar\ detector \cite{detector}, which operates at the 
PEP--II asymmetric--energy $e^+e^-$ collider at the Stanford Linear Accelerator Center
\cite{pep}. 
The sample
consists of 19.4\invfb\ taken at the $\Upsilon(4S)$ resonance,
corresponding to $21.3\times 10^6$
$e^+e^- \rightarrow \Upsilon(4S) \rightarrow B\overline B$ events. 
An additional sample of 2.2\invfb\ accumulated 
40\mev\ 
below the $\Upsilon(4S)$ resonance is used to estimate non--$B\overline B$ background.

Charge conjugation invariance is assumed for all channels quoted in this
paper, and the charge conjugate reactions are included in the analysis.
Quantities evaluated in the $\Upsilon(4S)$ rest frame are denoted by an asterisk; 
{\it e.g.}, $E^*_{b}$ is the energy of the $e^+$ and $e^-$ beams in the 
$\Upsilon(4S)$ rest frame.

The \babar\ detector, a general purpose solenoidal magnetic spectrometer, is 
described in detail elsewhere \cite{detector}. A silicon  vertex detector and a
cylindrical drift chamber in a 1.5-T solenoidal magnetic field are used to measure momenta and 
ionization energy loss of charged particles. 
Electrons and photons are identified by a 
CsI electromagnetic calorimeter (EMC). 

This analysis exploits in particular the information provided by the
EMC
consisting of 6580 CsI crystals, covering 90\% of 4$\pi$ 
in the $\Upsilon(4S)$ rest frame. 
The energy resolution has been measured directly with a radioactive source at low energy
and with electrons from Bhabha scattering at high energy.
The mass resolution of $\pi^0$ and $\eta$ candidates in which the two photons 
in the decay have approximately equal energy can be used to infer the energy resolution
at an energy less than 1\gev; the decay $\chi_{c1}\rightarrow J/\psi\gamma$ provides
an additional measurement at 500\mev. 
A fit to the energy dependence results in $\sigma_E/E = (2.3 \pm 0.3)\%/^4\sqrt{E} \, \oplus \, 
(1.9 \pm 0.1)\%$ \cite{detector}.

Energy deposits in the EMC are reconstructed
by grouping adjacent crystals with energy deposits greater than 
1\mev\ into {\it clusters}. Clusters with more than one local energy maximum are then split 
into {\it bumps}. 
The energy of each crystal is divided among the bumps by an iterative adjustment of the 
centers and energies of the bumps assuming electromagnetic shower shapes \cite{detector}. 
Next, all tracks reconstructed in the tracking volume are extrapolated to the 
EMC entrance and a track--bump matching probability is calculated for each pair.

All bumps with a matching probability smaller than $10^{-6}$ are treated as photon candidates.
Photons are selected  by requiring the bump shape to be compatible with an electromagnetic
shower, and by requiring the bump to have a minimum energy of 30\mev. 
In addition we accept only photon candidates which are isolated from any other bump in the event. 
This requirement selects against background from high--energy $\pi^0$ mesons, where the two 
photons from the decay of the $\pi^0$ meson strike the calorimeter in close proximity 
({\it merged $\pi^0$}).

The \babar\ detector is simulated by a {\tt GEANT}--based Monte Carlo procedure \cite{geant} that includes
beam--related background by mixing random trigger events into the Monte Carlo generated events.
The simulated events are processed in the same manner as data. 
The simulation is used to study background and optimize selection criteria, but
only enters the analysis directly through the calculation of the signal efficiency.

In order to select $B\overline B$ events, we require at least three tracks of good quality
in the event. The quality requirements for these tracks include a small impact parameter with respect to
the collision point along the beam direction (10\cm) and transverse to it (1.5\cm), 
a minimum number of 13 hits in the drift chamber and a momentum of 
$p<10\gevc$ in the laboratory frame. To help reject continuum background, the ratio of the 
second Fox--Wolfram moment to the zeroth Fox--Wolfram moment \cite{FoxWolfram} must be less than 0.9. 
We further require that there be two high--energy photon candidates with
an energy in the $\Upsilon(4S)$ rest frame between 1.5 and 3.5\gev. 
At this point, all remaining pairs of photons are considered candidates
for the decay $B^0 \rightarrow \gamma \gamma$.
If the event contains more than one such $B$ candidate all of them are kept for further analysis.

After this preselection additional requirements are imposed on the 
$B^0 \rightarrow \gamma \gamma$ candidates.
Photon bumps from the $B$ candidate must not contain noisy crystals or crystals which produce no
signals.
The second moment of the energy distribution around the cluster's centroid 
must be smaller than 0.002. This value has been optimized
to
reject 
remaining background from merged $\pi^0$ mesons.

Since $B$ mesons at the $\Upsilon(4S)$ resonance are produced nearly at rest, the decay 
$B^0 \rightarrow \gamma \gamma$ will contain two nearly back--to--back photons with 
$E^*_\gamma \approx 2.6 \gev$ in the $\Upsilon(4S)$ rest frame. This represents a clean signature and
makes this channel relatively easy to study experimentally. We  exploit this 
feature by considering only $B^0 \rightarrow \gamma \gamma$ candidates
which have at least one photon with $2.3 < E^*_\gamma < 3.0 \gev$.

In order to reject photons from $\pi^0$($\eta$) decays 
we combine each photon from the $B$ candidate with all the other 
photons in the event having energy greater than 50(250)\mev. The resulting $\pi^0$($\eta$)
candidates are required to have an invariant
mass beyond three standard deviations, or $3 \times 8.8 (18) \mevcc$, 
of the nominal $\pi^0$($\eta$) mass \cite{PDG}.


Reconstruction of exclusive final states from $B$ mesons produced at the \Y4S\ resonance benefits
from the beam energy constraint $E^*_B = E^*_{b}$. Thus, in the 
$\Upsilon(4S)$ rest frame the energies of the $B$ meson decay products must add up to the beam energy.
We calculate the energy difference $\Delta E \equiv E^*_{\gamma,1} + E^*_{\gamma,2} - E^*_{b}$
between the candidate $B^0$ meson and the beam energy 
in the \Y4S\ rest frame. The distribution of this quantity peaks at 0 GeV for true $B$ mesons,
and has a tail towards negative 
$\Delta E$ due to 
shower leakage in the EMC.
The resolution in $\Delta E$ is obtained from signal Monte Carlo events with a fit of the $\Delta E$ 
distribution to an empirical function \cite{CrystalBall} and
is $\sigma_{\Delta E} = 73\mev$.

The $B$ meson mass resolution is improved 
with the use of the 
beam energy constraint. 
We use the beam energy substituted mass
\mes\ $ \equiv \sqrt{E^{*2}_{b}-({\mathbf p}^*_{\gamma,1} + {\mathbf p}^*_{\gamma,2})^2}$.
The resolution on \mes\ is obtained from signal Monte Carlo events with a fit of the \mes\ distribution 
to an empirical function \cite{CrystalBall} and 
is $\sigma_{\mes} = 3.9\mevcc$.

For the purpose of determining numbers of events and efficiencies a rectangular 
signal region is defined. This region extends 2$\sigma$ 
in $\Delta E$ about 0\mev\ and 2$\sigma$ in \mes\ about the nominal mass $m_{B^0}$ of the $B^0$ meson.

The search for $B^0 \rightarrow \gamma \gamma$ was performed as a blind analysis by hiding
a 3$\sigma$ region in $\Delta E$ and \mes\ in on--resonance data until the development of 
the selection procedure was complete. 
This allows optimization of the selection and estimation of the background 
without the bias of knowing the number of events in the signal region.

Monte Carlo studies indicate that the main background arises from the process 
$e^+e^- \rightarrow q\overline q$ ($q = u$, $d$, $s$), referred to as continuum background
and modeled with the JETSET event generator \cite{JETSET}.
Such events 
exhibit a two--jet structure and contain
high momentum, approximately back--to--back tracks. 
One source of background includes photons
from initial-state radiation, others are photons from $\pi^0 \rightarrow \gamma \gamma$ 
and $\eta \rightarrow \gamma \gamma$ decays, where the decay is very asymmetric in the 
final-state photon energy.
Background from merged $\pi^0$ mesons is negligible. 

\begin{table}[tttt]
\caption{
Cumulative signal reconstruction efficiency as selection criteria are applied.
The first row shows the cumulative event selection efficiency. The additional rows give 
individual contributions to the $B$ candidate selection efficiency. The cumulative signal
reconstruction efficiency is 
given by the product of event selection and final $B$ reconstruction efficiency.
}
\label{tab:efficiency}
\begin{tabular}{lc}
  \hline
  \hline
  selection criteria& efficiency [\%] \\
  \hline
  cumulative event preselection             & 39.8   \\
  \hline
  photon energy $E^*_\gamma$                & 92.9   \\
  bump quality and second moment            & 86.8   \\
  $\pi^0$ and $\eta$ veto                   & 72.8   \\
  $|\cos(\theta_B)|$ and $|\cos(\theta_T)|$ & 40.1   \\
  signal region                             & 27.0   \\
  \hline 
  cumulative                                & 10.7   \\ 
  \hline
  \hline
\end{tabular}
\end{table}

To reduce continuum background, we calculate the angle $\theta^*_T$ between one of the photons
(chosen randomly) of the $B^0$ candidate and the thrust axis of the remaining tracks and neutral bumps 
in the event. The distribution of $|\cos\theta^*_T|$ is uniform for signal events and 
strongly peaked at 1 for continuum background events. 
We also calculate the angle $\theta^*_B$ between the momentum vector of the $B^0$ candidate 
and the beam axis in the $\Upsilon(4S)$ rest frame.
The distribution of $|\cos\theta^*_B|$ is uniform
for continuum background and  follows a $\sin^2\theta^*_B$ distribution for signal events.
The requirements for both $|\cos\theta^*_T|$ and $|\cos\theta^*_B|$ have been 
optimized to maximize the statistical significance $N_S/\sqrt{N_S+N_B}$, where $N_S$ is the number of 
signal candidates expected assuming for the branching fraction 
${\cal B}(B^0 \rightarrow \gamma \gamma ) = 1 \times 10^{-8}$ \cite{b0gammagammaSUSY}
and $N_B$ is the expected number of background candidates determined from continuum 
Monte Carlo simulation and  off--resonance data. 
We require $|\cos\theta^*_T|<0.57$ and $|\cos\theta^*_B|<0.81$.
If more than one $B$ meson candidate per event remains after this selection, which occurs in 
less than 0.1\% of the events analyzed, we select the candidate with the smallest $|\Delta E|$.
After all these selection criteria
the overall efficiency for $B^0 \rightarrow \gamma \gamma$ decays is determined from 
the Monte Carlo simulation to be $( 10.7\pm 0.2 )\%$, where the error is purely statistical.
Table \ref{tab:efficiency} shows the cumulative signal reconstruction efficiency as the 
selection criteria are applied.

\begin{figure}[ttt]
    \epsfig{file=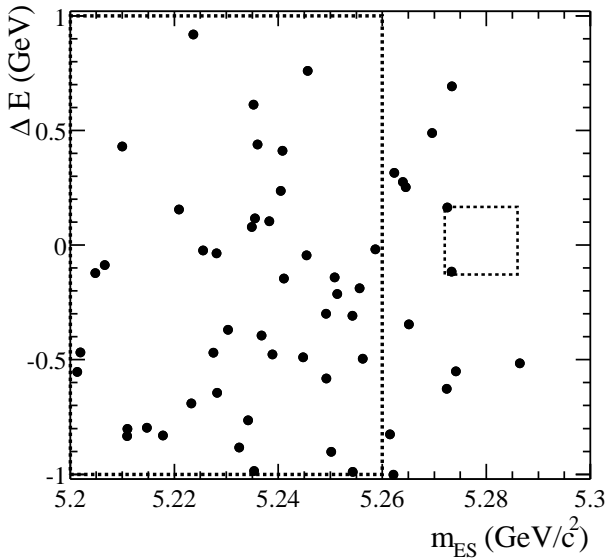,width=\linewidth}   
    \caption{
      Energy difference 
      $\Delta E$
      between the candidate $B^0$ meson and the beam energy 
      in the \Y4S\ rest frame versus
      beam energy substituted mass
      \mes\ 
      for on--resonance data. 
      We observe one event in the signal region, outlined as a black dashed box about $\Delta E = 0$\gev,
      consistent with the expected background. The dashed box on the left shows the sideband used
      for background estimation.
      \label{fig:final}
     }
\end{figure}

\begin{figure}[ttt]
\epsfig{file=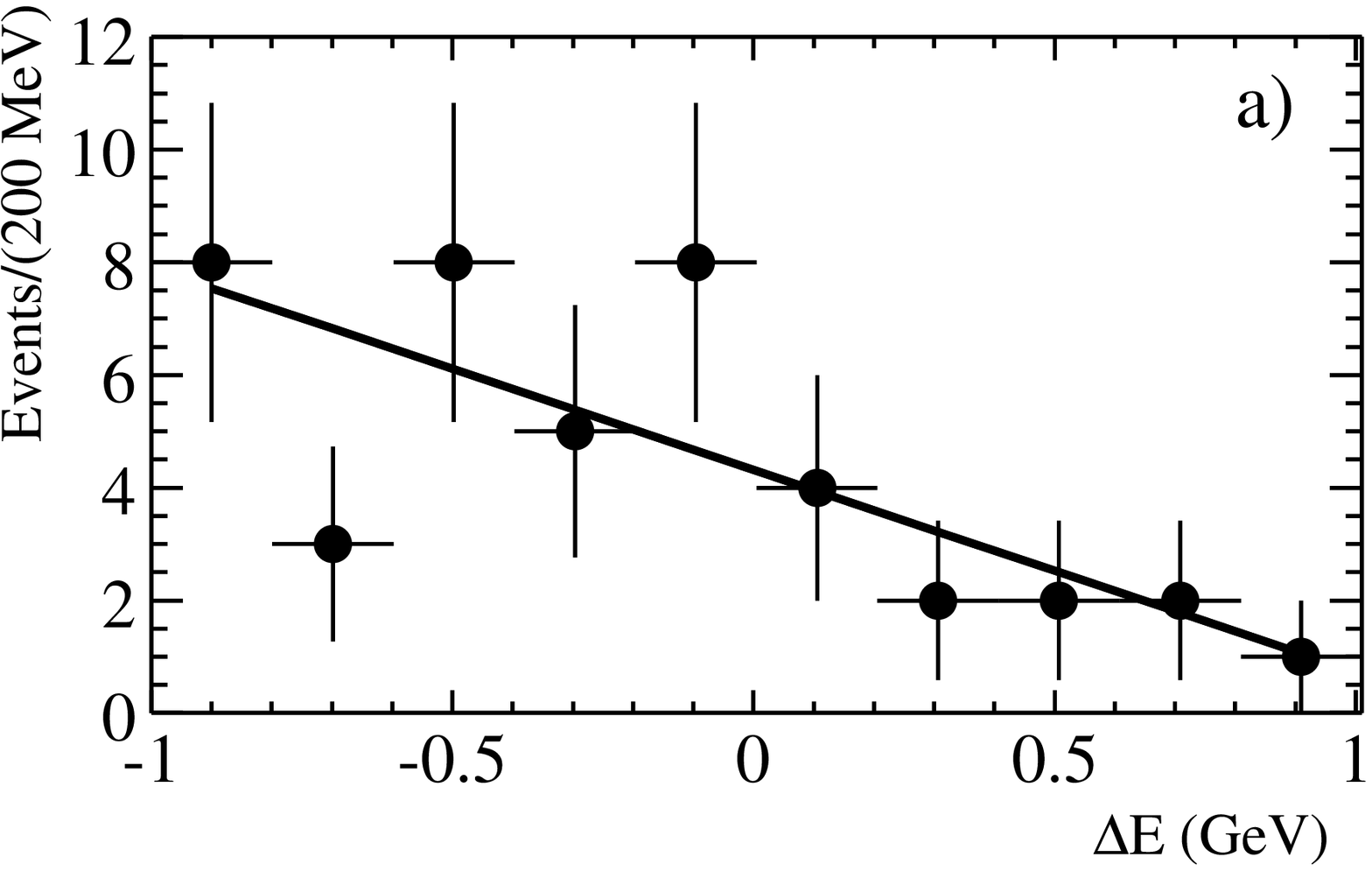,width=\linewidth}
\epsfig{file=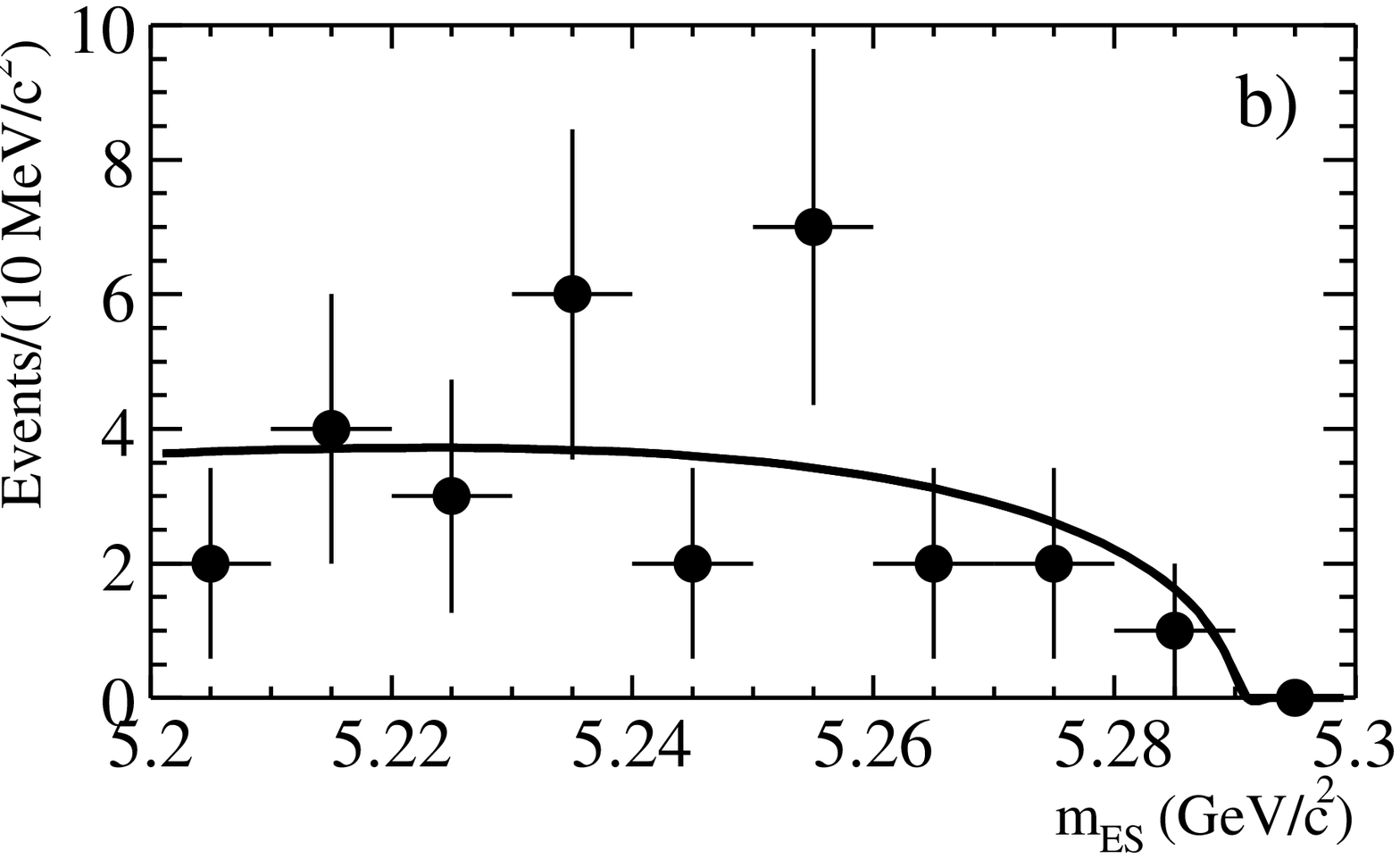,width=\linewidth}
\caption{
  a) Fit to the $\Delta E$ distribution in the grand sideband
  to a first-order polynomial; b) fit of the \mes\ distribution in the lower 
  sideband 
  with the ARGUS function \cite{Argus}. See text for the definition of the
  sidebands.
  \label{fig:background}}
\end{figure}

A single event in the on--resonance data meets these selection criteria, as shown in Fig.~\ref{fig:final}.
A number of 
exclusive decay modes that can mimic $B^0 \rightarrow \gamma \gamma$ decays have been studied
with high statistics (equivalent to $1.2-1.7 \times 10^4\invfb$ assuming branching fractions 
of the order of
10$^{-6}$). We expect negligible contributions from  
$B^0 \rightarrow \eta\eta$, $K^{*0}\gamma$, $\rho^{0}\gamma$, and $\pi^0\pi^0$, and a combined 
contribution of $0.7\times 10^{-3}$ events from $B^\pm \rightarrow \rho^\pm(\pi^\pm\pi^0)\gamma$ and 
$B^0 \rightarrow \omega(\pi^0\gamma)\gamma$.
To further explore the question of remaining background in the signal region, 
we define the {\it grand sideband} 
consisting of a rectangular region within the limits 
$-1.0 < \Delta E < 1.0 \gev$ and $5.20 < \mes < 5.26 \gevcc$ 
(see Fig. \ref{fig:final}, left dashed box).
In this region we find a prediction of $34\pm 9$ events from continuum 
Monte Carlo simulations, in good agreement
with the observation of $43\pm 7$ ($44\pm 20$) events from on--resonance data (off-resonance data
of 2.2 fb$^{-1}$ scaled to the full analyzed luminosity of 19.4 fb$^{-1}$).
We parameterize the background using on--resonance data.
The background in  $\Delta E$ is parameterized in the grand sideband 
with a first-order polynomial (see Fig. \ref{fig:background}a); the background in \mes
is parameterized in the {\it lower sideband}, which is a rectangular region within the limits 
$-1.0 < \Delta E < -0.2 \gev$ and $5.20 < \mes < 5.29 \gevcc$, with an empirical threshold function
first employed by the ARGUS collaboration \cite{Argus} (see Fig. \ref{fig:background}b).
Both parameterizations describe the corresponding distribution very well with a $\chi^2$, normalized
to the number of degrees of freedom, of about 0.8.
Using this parameterization we are able to extrapolate the on--resonance grand sideband data 
into the signal region and find an expectation of $0.9^{+0.4}_{-0.3}$ events. 
This is consistent 
with the hypothesis that the observed event in the signal region is due to continuum background.
Nevertheless, 
we choose to 
quote a conservative upper limit, assuming that the observed event in the signal region is 
in fact due to the decay $B^0 \rightarrow \gamma \gamma$. 
We use Poisson statistics to set an upper limit on the branching fraction.
The upper limit on the branching fraction $\cal B$ is obtained from 
${\cal B} = N_{UL}/(\epsilon \cdot (N_{\Bz}+ N_{\Bzb}))$,
where $N_{UL}$ is the upper limit on the number of observed events, $\epsilon$ 
the signal reconstruction efficiency of (10.7 $\pm$ 0.2)\% and 
$N_{\Bz}+ N_{\Bzb}$ is the number of produced \Bz and \Bzb mesons. 
$N_{\Bz}+ N_{\Bzb}$ is equal to the number
of $\Upsilon(4S)$ events since we assume the number of \Bz\Bzb events to be 50 \% of the number
of produced $\Upsilon(4S)$ events.
This yields an upper limit on the branching fraction, based on statistics alone,
of ${\cal B}(B^0 \rightarrow \gamma \gamma ) < 1.7 \times 10^{-6}$
at the 90\% confidence level.

\begin{table}[ttt]
\caption
{Summary of systematic uncertainties on the signal efficiency and the number
of produced \FourS\ as an error on the branching fraction determination.
The total systematic uncertainty is the sum of the individual contributions added
in quadrature.}
\label{tab:systematics}
  \begin{tabular}{ l r }
    \hline
    \hline
      systematic uncertainty                   & ($\Delta${$\cal B$}/{$\cal B$})\% \\
    \hline
    Number of produced \FourS\           &  1.6 \\
    Photon detection efficiency          &  6.5 \\ 
    $\eta$ veto                          &  2.0 \\
    $\pi^0$ veto                         &  2.0 \\
    $\Delta E$ selection                 &  5.3 \\
    \mes\ selection                      &  2.6 \\
    Track finding efficiency             &  1.8 \\ 
    Number of signal Monte Carlo events  &  2.0 \\
    \hline
    Total                                &  9.6 \\
    \hline
    \hline
  \end{tabular}
\end{table}
Systematic effects arise from the modeling of the signal efficiency and the estimation 
of the number of $B$ mesons in the data sample.
A summary of all systematic errors
is provided in Table~\ref{tab:systematics}.
The most significant sources are
the photon detection efficiency 
and 
the $\Delta E$ selection 
due to the
uncertainty in the photon energy scale and photon energy resolution.
The systematic uncertainty on the photon detection efficiency has been 
determined from a study which compares the precisely know ratio \cite{PDG} of 
the $\tau \rightarrow \pi\pi^0\nu_\tau$ and $\tau \rightarrow \pi\pi^0\pi^0\nu_\tau$ rate
in Monte Carlo events and data. 
This uncertainty depends on the event multiplicity, whose effect is estimated by embedding 
photon bumps from radiative Bhabha events into both generic $B$ meson
and generic $B$ meson Monte Carlo events.
The uncertainty in the energy scale is estimated with a study of symmetric 
$\eta \rightarrow \gamma \gamma$ decays,
where both photons are within a narrow energy range. 
Systematic shifts of the reconstructed 
$\eta$ mass from the nominal value measure the uncertainty in the energy scale in this 
energy range. 

In order to include our systematic uncertainty in the determination of the upper limit,
we follow a prescription given by \cite{Cousins}.
The branching fraction $\cal B$ is calculated as  ${\cal B} = n / S$, where $n$ is the number of 
observed events and $S = 2.3\times 10^6$ is the sensitivity, 
given by the product of the number of $B^0$\Bzb\ 
events 
and the overall $B^0 \rightarrow \gamma \gamma$ selection efficiency.
Assuming a normal distribution for the uncertainty in $1/S$,
the systematic uncertainty is accounted for by convoluting the Poisson probability
distribution for the assumed branching fraction with a Gaussian error distribution for $1/S$.
Our total systematic uncertainty of 9.6\% included in this way has a negligible effect on the upper limit.


In summary, 
we have performed a search for the decay $B^0 \rightarrow \gamma \gamma$. We observe one event in 
the signal region and infer an upper limit on the branching fraction
of $${\cal B}(B^0 \rightarrow \gamma \gamma ) < 1.7 \times 10^{-6}$$
at the 90\% confidence level.
This result improves the existing limit 
\cite{L3} 
by over a factor of 20.

We are grateful for the 
extraordinary contributions of our \pep2\ colleagues in
achieving the excellent luminosity and machine conditions
that have made this work possible.
The collaborating institutions wish to thank 
SLAC for its support and the kind hospitality extended to them. 
This work is supported by
DOE
and NSF (USA),
NSERC (Canada),
IHEP (China),
CEA and
CNRS-IN2P3
(France),
BMBF
(Germany),
INFN (Italy),
NFR (Norway),
MIST (Russia), and
PPARC (United Kingdom). 
Individuals have received support from the Swiss NSF, 
A.~P.~Sloan Foundation, 
Research Corporation,
and Alexander von Humboldt Foundation.

\end{document}